\documentclass[reprint,aps,pra,showpacs,amsmath,amssymb,floatfix,nofootinbib]{revtex4-1}
\usepackage{graphicx,times}
\usepackage[bookmarks=true]{hyperref}

\newcommand{\Span}{\mathrm{span}}
\newcommand{\rank}{\mathrm{rank}}
\newcommand{\ket}[1]{|#1\rangle}
\newcommand{\bra}[1]{\langle#1|}
\newcommand{\braket}[1]{\langle #1 \rangle}
\newcommand{\ketbra}[2]{\ket{#1}\!\bra{#2}}
\newcommand{\trnsp}[1]{#1^{\mathrm{t}}}

\newcommand{\ii}{\mathrm{i}}
\newcommand{\ee}{\mathrm{e}}
\newcommand{\cR}{\mathcal{R}}

\begin{document}

\title{Scheme for measurement-based quantum computation on 
projected entangled-pair states}

\author{Mear M. R. Koochakie}
\email {koochakie@gmail.com}
\affiliation{Department of Physics,
Sharif University of Technology, Tehran, Iran}


\begin{abstract}
Recently it has been shown that projected entangled-pair states can be considered as a (physically motivated) resource state for measurement-based quantum computation. Here we elaborate on how to construct a deterministic measurement-based quantum computation model on projected entangled-pair states. In particular, we use this scheme to build a 4-level many-body model for universal quantum computation on qubits. We also derive a 3-local parent Hamiltonian for this model, whose ground state is non-degenerate.
\end{abstract}


\maketitle

\section{Introduction}
Quantum computation 
has shown promising features which apparently have no counterparts in classical computing models. The two prominent examples of such features are in Shor's quantum factoring algorithm \cite{shor94} and Grover's quantum search algorithm \cite{grover}.

Despite its power, quantum computation is challenging to implement \cite{div}. The traditional quantum circuit model \cite{circuit}, which is used to express most quantum algorithms, needs a practical implementation of one and two-qubit gates, which  seems difficult in most existing realizations \cite{naka-art}. To address this problem, various schemes, notably measurement-based quantum computation (MQC) \cite{rass, nature}, have been proposed. MQC is as powerful as quantum circuit model, but based on a different notion of required operations for implementation. In this model one should first prepare a suitable highly entangled state and then apply an adaptive single-site measurement pattern on it. The result of the computation will be deduced from the measurement outcomes. 

The specific resource state of the MQC model, introduced in Ref.~\cite{rass}, is the cluster state. 
However, there is also a vast literature on a variety of MQC models with different resource states \cite{rud,gross07,gross07pra,Brennen08,chen,cai,wei11,li}, which have been argued to offer more physical base for MQC rather than the cluster state.
These works were mostly influenced by the seminal study of Ref.~\cite{vc04}, in which a relation between the cluster state and valence-bond states was suggested, and modeled measurement-based quantum computation by teleportation-based quantum computation on valence-bond states.

It should be noted that most of the recent works along this line fall into the category of ``random-length'' MQC models, in the sense that the number of basic steps  required to 
perform basic elements of a computation is \emph{a priori} unknown. In this respect such random-length MQC can be viewed as a sort of \emph{stochastic} MQC models \cite{MoBri10} with close to  one success probability. 
Despite the appeal of such stochastic models, they may be difficult to be employed in design and analysis of quantum algorithms due to the very lack of the knowledge of length of each step.  
A natural difficulty of this type is exemplified in, e.g.,  synchronization of two distinct threads of computation in order for implementing a two-qubit gate. Such an operation, although theoretically possible, may require a considerable overhead resources to take care of the randomness.
To avoid these difficulties, here we restrict our attention to  \emph{deterministic} MQC. 
 Note that by ``deterministic'' we mean that the computation can be performed in a pre-determined number of steps giving desired result with certainty.
 
Further generalization of MQC models has been introduced by replacing projective measurement with generalized one, see, e.g., Refs. \cite{nest06,cai09}.
According to, e.g., Naimark's theorem \cite{nai43}, the implementation of a generalized measurement is reducible to addition of an ancillary site, implementation of a two-body gate and a projective measurement on the ancilla. 
For our purpose in this work, however, such a generalization is not necessary, given the fact that in most of the existing quantum circuit models the only prohibitive barrier for implementation is inaccessibility of a reliable two-gubit gate.
Bearing this in mind, we prefer to restrict ourselves to projective measurements for our investigation here.



Here following the proposition of Refs.~\cite{gross07,gross07pra}, we consider our resource state to be ``Projected Entangled-Pair States'' (PEPS), which have been used to represent ground state of an important class of two-dimensional quantum systems \cite{Cir06}.
Specifically, we elaborate on a more systematic approach toward a PEPS-MQC in which the by-product(error) propagation through computation partly determines the structure of the resource state and the measurement pattern. 

This approach is illustrated by devising a particular PEPS-MQC model. This model is constituted from four-level sites on a honeycomb lattice, through which qubit quantum computation is performed. 
 A characteristic of this model is that, in comparison to the one-way model if require less number of sites to perform a single-qubit gate. 
 As such, in computational tasks in which majority of gates are single-qubit operation, this model can outperform the one-way model in the number of necessary sites.
  We also provide a three-local parent Hamiltonian for this model.
  
\emph{Notation}.---Throughout this article the following assumptions are made:
\begin{enumerate}
  \item Let $M$ be a list of rank-$n$, then we denotes its
    elements by $M(i_1,i_2,\ldots,i_n)$, where $i_j$ are non-negative integers.
    E.g., 
    $ E\,=\,\big(E(0),E(1),\ldots,E(d-1)\big) $
    is a $d$-dimensional rank-1 list. The elements of a list could
    be any type of linear objects (numbers, vectors, matrices, tensors, and so on).


  \item We assumed that quantum information evolves from right towards
  left. Also index of columns increases from right to left (\ldots, 3,
  2, 1). 

  \item We assume that $d$ possible output values of a measurement are
  in the set $\{0,1,\ldots,d-1\}$. 

  \item  $X$ and $Z$ are $\sigma_x$ and $\sigma_z$ Pauli matrices.

\end{enumerate}

\section{MQC on PEPS}
The goal of each MQC model, by introducing a state and a pattern of
adaptive measurements on this state, is to accomplish a quantum computation. 
We should first define our resource state, on which MQC is implemented, and specify by what type of measurements we can simulate the circuit model on it. 
According to Ref.~\cite{gross07}, the one-way model \cite{rass} can be considered as a
computation on the \emph{correlation space} of a specific MPS which is realized
by adaptive single-site measurements of the MPS. 
But this one-dimensional state can only generates $\mathrm{U}(2)$
elements. 
Thus by adding a second dimension (and indeed working on a
PEPS) one can also simulate two-qubit gates.

An MPS is a state, denoted by $\ket{\mathrm{MPS}}$, that can be written in the standard basis as
\begin{multline}\label{eq:mps}
\ket{\mathrm{MPS}} = \\
\sum_{i_1,i_2,\dotsc,i_N} \bra{L} A_N(i_N)\dotsm A_2(i_2)A_1(i_1) \ket{R}\:
\ket{i_N,\dotsc,i_2,i_1},
\end{multline}
in which $N$ is the number of sites, and $0 \leq i < d$ (i.e., each site
has $d$ levels) and $A_j(i)$ is a $D\times D$ matrix which attributed to
 the state $\ket{i}$ of site $j$. 
$\bra{L}$ and $\ket{R}$ are two boundary vectors. 
 As can be seen in Eq.~\eqref{eq:mps}, the amplitude coefficients of the MPS are computed by a set of matrix multiplications in the so-called \emph{correlation space}. 
Note that the dimension of the correlation space, $D$, can differ from the dimension of the physical sites, $d$ (number of levels).

\subsection{One quDit-gates}
Let us now clarify the idea introduced in Refs.~\cite{gross07,gross07pra} for single-quDit gates. 
Assume we have an MPS with a list of matrices, $A$. 
If one measures the $j$th site of the chain and the resulting outcome state is $\ket{\varphi}$, then the state changes to (up to a normalization)
\begin{multline}\label{eq:mps_phi}
 \ket{\varphi}_j\bra{\varphi} \cdot
\ket{\mathrm{MPS}}\; =\\ \sum_{i_1,\dotsc,i_{j-1},i_{j+1},\dotsc,i_N}
\bra{L} A_N(i_N)\dotsm A_j[\varphi] \dotsm A_1(i_1) \ket{R}\\
\ket{i_N,\dotsc ,i_{j+1}}\ket{\varphi}\ket{i_{j-1},\dotsc ,i_1},
\end{multline}
where $A_j[\varphi]:= \sum_i\varphi^\ast(i) A_j(i)$ and $\ket{\varphi}
=\sum_i \varphi(i)\ket{i}$. As can be seen, if one can control the outcome of a measurement at site $j$, any operator in $\Span( A_j )\equiv \{\sum_i \alpha_i A_j(i)\}$ (up to some normalization) can be realized at the correlation space by a single-site measurement.

\begin{figure}
\centering
  \includegraphics[width=87mm]{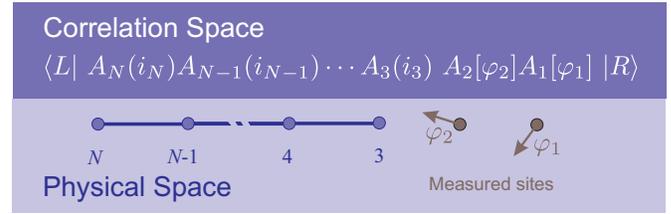}\\
  \caption{MQC on MPS: Measurement of the physical sites realizes the (one-quDit) quantum computation in the correlation space.}\label{fig:mqc-on-mps}
\end{figure}

Now consider we measure $\ell$ sequential sites after site $k$ of the MPS, so that in
the correlation space we have the following product:
\begin{equation} \label{eq:prod-U}
U=A_{k+\ell}[\varphi_\ell]\dotsm A_{k+2}[\varphi_2]A_{k+1}[\varphi_1], 
\end{equation}
 where $\ket{\varphi_j}$ is the $j$th measurement outcome state. Hence, if we choose
suitable $A_j$ as lists of matrices and suitable measurement bases,
one can construct all $U\in \mathrm{U}(D)$ in the correlation space. 
In other words, one achieves an arbitrary single-quDit gate in the correlation space (Fig.~\ref{fig:mqc-on-mps}).
This sums up our short review of the proposition of Ref.~\cite{gross07}, except the proposal that circumvents the side effects of the measurements.
%

The above proposal clearly necessitates a ``universal'' list of matrices to implement any arbitrary one-quDit gate. 
It is intriguing then to see what MPSs may have such properties, whose associated $A$ constitutes a universal list.
This question maybe difficult to answer in general. 
However, fortunately this question can be answered in some special cases,
 e.g., Ref.~\cite{gross-web} discusses the case of $D=d=2$ (qubit computation on 2-level sites) on translationally invariant MPSs ($A_j=A$) -- computational quantum wires.
Further discussion of the above question, however, is not the aim of our paper.
\subsection{By-product}
Measurement outcomes occur randomly in quantum mechanics. 
Thus it may happen that rather than a desired $\ket{\varphi}$ one obtains a different state $\ket{\varphi'}$ at the physical site. 
In the scenario where one uses MQC for deterministic quantum information processing, it is crucial to not lose any information.
 This fact constrains the possible matrix list $A_j$ and the set of possible measurements on the corresponding site in such a way that $A_j[\varphi]$ should be always invertible for each one of possible outcome states $\ket{\varphi}$ of an acceptable measurement. 
 This requirement, although not necessary for quantum computation,  we adopt here, since 
  it facilitates the simulation of circuit model quantum computation in the correlation space.
  
  Let us assumes that single-quDit gate $U$ entails measurement result states $\ket{\varphi_\ell}\cdots \ket{\varphi_2}\ket{\varphi_1}$, but instead we obtain  $\ket{\varphi'_\ell}\cdots \ket{\varphi'_2}\ket{\varphi'_1}$.
  Here $\ket{\varphi'_i}$ may or may not be equal to the desired state $\ket{\varphi_i}$. 
  The overall computation then can be interpreted as performing 
\begin{equation}\label{eq:err}
O = E\,U,
\end{equation}
where $O := A_{k+\ell}[\varphi'_\ell]\dotsm A_{k+2}[\varphi'_2]A_{k+1}[\varphi'_1]$ and $E=O U^\dagger$ is indeed an ``by-product operator'', whose invertibility is carried over from  $A_{k+i}[\varphi'_i]$s.
Such by-products are, in general, unavoidable in measurements, but need to be circumvented or compensated somehow.
Reference~\cite{gross07} deals with this problem by the ``trial until success'' approach, which in turn brings up a random-length MQC model.
In deterministic MQC, however, this randomness should be avoided, e.g., by unambiguously tracking the by-product propagation through the model and compensating these by-products in a deterministic manner.
 
To provide a model which is able to perform universal quantum computation, one needs to also consider implementation of two-quDit gates in the correlation space. 
 In fact, how the measurement by-products would pass through these gates puts stringent conditions on the applicability of the computation model in face of measurement by-products.
  These conditions together with invertibility of $A_i[\varphi]$s and existence of a universal $A$-list characterize our MQC-PEPS computation model.
\subsection{Readout}
This is essential to recall that all the computation thus far has been performed in the (virtual) correlation space. 
However, at the end of the computation one needs to realize the resulting state $\ket{\Phi}$ (e.g., $\ket{\Phi}=U\ket{R}$ in one-dimensional case) in the physical space on real sites. 
In the following, we elaborate more systematically on the readout process (See Ref.~\cite{gross07} for an example.)

Consider an $N$-site spin chain whose sites except the last one  have been measured. 
The resulting physical state $\ket{\Psi}$ becomes
\begin{equation} \label{eq:readout}
  \alpha \sum_i \braket{L|A_N(i)|\Phi}
\ket{i}_N := \sum_i \braket{L(i)|\Phi} \ket{i}_N,
\end{equation}
where $\alpha$ is a normalization factor.
Alternatively, one can think of the extraction process as the operation of a linear map
\begin{equation}
\cR := \alpha \sum_i \ket{i}\bra{L(i)}
\end{equation}
 on the correlation space $\ket{\Phi}$. $\cR$ is a $d\times D$ matrix, which needs to have the property $\rank(\cR) = D$ if it is supposed not to lose any information.
$\cR$ is also required to preserve the orthogonality of input vectors, which is satisfied if its columns are chosen from a $d \times d$ unitary matrix.
In the case the correlation space state $\ket{\Phi}$ carries a by-product $E$ the extracted state would become 
\begin{equation}\label{eq:readout-operator}
\ket{\Psi} = \cR E\, \ket{\Phi},
\end{equation}
in which $E$ needs to be unitary in order for orthogonality preserving.

 The above requirements for the readout process can be relaxed in random-length MQC, where the universality of MPS is argued to be sufficient for the extraction of states from the correlation space \cite{cai09}.
 \subsection{Two-quDit gate}
 In addition to the single-quDit gates it is necessary to find an entangling gate ($W$) between two
  quDits.
 A required constraint for this two-quDit gate in our model is that it needs to leave uncorrelated local by-products local.
   This property puts limiting conditions on the two-quDit gate given a set of by-products, or vice versa.
Thus we shall need only single-site by-product correction operations for an arbitrary computation.
  
 We stress again that it is the propagation of by-products through the computation which characterizes our specific computational model in this paper.
 A natural requirement to keep locality of the by-products after passing through the entangling gate $W$ is that for any pair of acceptable by-products $E$ and $F$ we have 
 \begin{equation}\label{eq:by-product-prop}
 W\, E \otimes F = G\otimes H\,W,
 \end{equation} 
where $G,\,H \in \mathrm{U}(D)$ are not necessarily members of acceptable by-products.
Solving this equation for a given $W$ (to obtain the set of acceptable by-products) is in general difficult. 
In Appendix~\ref{seq:err-two-qubit}, we provide a systematic solution for Eq.~\eqref{eq:by-product-prop} in the case of $D=2$ (qubits).
For example, all local gates that pass through the CZ gate locally can be parametrized as
\begin{equation}\label{eq:local-pass-CZ}
L(\theta_1,\theta_2,i,j)= Z(\theta_1) \Sigma(i) \otimes Z(\theta_2)\Sigma(j),
\end{equation}
where $\Sigma := \left( \openone,\, X,\, Y,\, Z\right)$, and $Z(\theta):=\mathrm{exp}\left(\frac{\ii}{2}\theta\, Z \right)$, which is an arbitrary rotation around the z-axis.
Therefore, the list of acceptable by-products for the CZ gate is
\begin{equation}\label{eq:CZ-list-err}
E_{\mathrm{CZ}}(\theta,i) = Z(\theta)\Sigma(i). 
\end{equation}
Choosing another gate, e.g.,  $\mathrm{exp}\left( \frac{\ii}{2} \gamma\, Z\otimes Z \right)$ ($\gamma = \tfrac{\pi}{2}$ corresponds to a local equivalent of the CZ gate) changes the set of acceptable local by-products to (see Appendix~\ref{seq:err-two-qubit})
\begin{equation}
E_{\gamma}(\theta)=Z(\theta)\, \sigma\ \mathrm{if}\ \gamma \neq \frac{\pi}{2},
\end{equation} 
where $\sigma$ is a specific Pauli matrix or the identity matrix for list $E$.
Thus it is evident that the choice of entangling operator puts intimate restrictions on the set of acceptable local by-products.
  
Here we lay out a systematic approach for how to realize a two-quDit gate over PEPS (see Ref.~\cite{gross07} for an example). 
Adding an extra dimension, hence the very necessity of PEPS rather than MPS, is justified because of the fact that one cannot implement a two-quDit operation only through multiplication of the operators of an MPS.

\begin{figure}
 \centering
   \includegraphics[height=36mm]{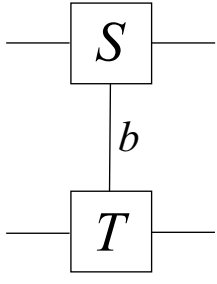}\\
   \caption{Vertical PEPS structure.}\label{fig:two-quDit-PEPS}
 \end{figure}

 Figure~\ref{fig:two-quDit-PEPS} illustrates two ``up'' and ``down'' sites of a two-dimensional lattice, whose corresponding tensors associated to level $i$ and level $j$ of these sites are denoted by, respectively, $S(i)$ and $T(j)$.
 The collection of these two sites can be seen as a rank-2 list of tensors
 \begin{equation}
 \mathcal{C}\{ST\} (i,j) = \mathcal{C} \{S(i)T(j)\},
 \end{equation}
 where the contraction $\mathcal{C}$ here means
 \begin{equation}
\left(  \mathcal{C}\{S(i)T(j)\} \right)_{mn,kl} = \sum_b S(i)_{bmk} T(j)_{bnl}.
 \end{equation}
 The tensors $S(i)$ and $S(j)$ can be decomposed as follows:
 \begin{align}\label{eq:expand-tensors}
 S(i) &= \sum_{b,\kappa} M(i)_{b,\kappa}\, \ket{b}\otimes B_{\kappa} \\
 	&= M(i)\otimes \openone \sum_{\kappa} \ket{\kappa}\otimes B_{\kappa}, \\
 T(j) &= \sum_{b,\lambda} N(j)_{b,\lambda}\, \ket{b}\otimes B_{\lambda} \\
 	&= N(j) \otimes \openone \sum_{\lambda} \ket{\lambda} \otimes B_{\lambda},
 \end{align}
 where $\{\ket{b}\}$ constitutes a basis for the $D'$-dimensional vertical correlation space,  $\{B_{\lambda}\}$ is a basis for space of $D\times D$ matrices, and $M\,(N)$ is a list of $D'\times D^2$ matrices.
 Hence
 \begin{equation}
 \mathcal{C}\{ ST \}(i,j) = \sum_{\kappa,\lambda} \left( \trnsp{N}(j) M(i) \right)_{\lambda,\kappa} \, B_{\kappa} \otimes B_{\lambda},
 \end{equation}
 in which superscript `t' denotes transposition.
 Now suppose one measures both sites separately and the resulting outcome states are $\ket{\varphi}$ and $\ket{\psi}$.
 In the correlation space this, it yields
 \begin{equation}\label{eq:2qubitgate}
 \mathcal{C}\{ST\} [\varphi,\psi] = \sum_{\kappa,\lambda} \left( \trnsp{N}[\psi] M[\varphi] \right)_{\lambda,\kappa} \, B_{\kappa} \otimes B_{\lambda},
 \end{equation}
 which we required to be the sought after entangling gate $W$.
As discussed earlier in Eq.~\eqref{eq:by-product-prop}, when the measurement result states are $\ket{\varphi'}$ and $\ket{\psi'}$ (rather than $\ket{\varphi}$ and $\ket{\psi}$) it is required that 
 \begin{equation}\label{eq:by-product-twoquDit-peps}
\mathcal{C}\{ST\} [\varphi',\psi'] = G\otimes H\, W.
 \end{equation}
 This is the very condition of locality of the by-products.
 Equation~\eqref{eq:by-product-twoquDit-peps} implies
 \begin{equation}\label{eqs:loc-err-two-quDit}
 \trnsp{N}[\psi']M[\varphi']=  \trnsp{\left(N[\psi]h\right)} M[\varphi]\, g,
   \end{equation}
 where $g$ and $h$ are $D^2 \times D^2$ matrices satisfying
 \begin{equation}
 GB_\mu = \sum_\nu g_{\mu \nu} B_\nu,\quad
 HB_\mu = \sum_\nu h_{\mu \nu} B_\nu.
 \end{equation}
Existence of a solution for Eq.~\eqref{eqs:loc-err-two-quDit} for $g$ and $h$ is necessary to   have local by-products, although this is not sufficient since not each $g(h)$ corresponds to a $G(H)$.
 
The criteria we derived in this section can guide people to design deterministic PEPS-based MQC models following recipe below:

\begin{itemize}
\item Select a two-quDit gate,
\item find all local unitary operators crossing the two-quDit gate locally (the method of Appendix~\ref{seq:err-two-qubit} for the qubit case),
\item choose the tensors and measurement bases such that, in addition to the universality, by-products of measurements belong to the above set of unitaries.
\end{itemize}
 
\section{A qubit MQC model on four-level honeycomb lattice}\label{sec:model}
\begin{figure}
\centering
  \includegraphics[width=40mm]{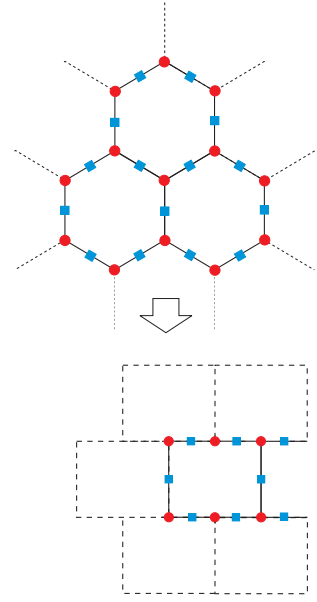}\\
  \caption{Honeycomb lattice. The horizontal (vertical) edges are used to perform single-qubit (two-qubit) gates.}\label{fig:model}
\end{figure}

In this section, we introduce a model for qubit MQC in correlation space of a 4-level honeycomb lattice.
This lattice includes two types of sites: vertex sites (circles) and edge sites (squares) (Fig.~\ref{fig:model}). 
Each site corresponds to a 4-level system. 
We should introduce four tensors for each type of sites. 
To the square sites, we assign the following list of $2\times 2$ matrices:
\begin{equation}
A=\tfrac{1}{\sqrt{2}}\big( \begin{bmatrix}
  1 &  0\\
0 & 1
\end{bmatrix}, 
\begin{bmatrix}
  0 &  1\\
\ii & 0
\end{bmatrix}, 
\begin{bmatrix}
  0 &  \ii \\
1 & 0
\end{bmatrix},
\begin{bmatrix}
  1 &  0\\
0 & -1
\end{bmatrix} \big). \label{eq:model-A}
\end{equation} 
The elements of $A$ constitute an orthonormal basis for $2\times 2$ matrices,
in the sense of the Hilbert-Schmidt inner product $\langle X, Y \rangle := \mathrm{Tr} [X^\dagger Y]$; i.e. , $\langle A(i),A(j) \rangle = \delta_{ij}$. 
The fact that $A$ is a basis matrices can simplify various steps of the computation scenario in the model.
For example, one needs only one measurement to perform any one-qubit gate $U$,
since one can always find a measurement outcome state  $\ket{\varphi(0)}$ such that 
$U = A[\varphi(0)]=\sum_i \varphi_i^\ast(0) A(i)$. 
We also need to have three more orthonormal states $\ket{\varphi(i)} \,(i=1,2,3)$ for the measurement basis.
This orthonormality yields
\begin{equation}
\langle A[\varphi(i)], A[\varphi(j)] \rangle = \delta_{ij},
\end{equation}
and
\begin{equation}
\langle E(i), E(j) \rangle = \delta_{ij},
\end{equation}
if we define the by-products through $A[\varphi(i)]=E(i)U$.
Thus, the by-products also become orthonormal.

As the entangling two-qubit gate, we consider the CZ gate.
The list of the acceptable by-products then is evident from Eq.~\eqref{eq:CZ-list-err} as
$E(i)=Z(\theta)\Sigma(i)$
for a given $\theta$.
Assuming $E(0)=\openone$ implies that 
\begin{equation}
E=\Sigma,
\end{equation}
 i.e., the list of the Pauli matrices extended with the unity matrix.

\subsection{One-qubit gates}
Any unitary $ U \in \mathrm{SU}(2)$ can be written as
\begin{equation}
U = \begin{bmatrix}
a & b \\
-b^\ast & a^\ast
\end{bmatrix},
\end{equation}
where $|a|^2+|b|^2=1$.
One can realize the operator $U$ (up to some Pauli by-product) in the correlation space by choosing the measurement basis according to Table~\ref{tb:1qubitgate}.
\begin{table}[bp]
\caption{Measurement basis and achieved single-qubit gates. The first column contains the measurement outcome states, and the next column shows the corresponding achieved gates at the correlation space.} 
\begin{tabular}{lr}
\hline \hline
Measurement basis & Achieved gate \\
\hline
    $\left(a+a^\ast, b^\ast - \ii b, \ii b^\ast - b, a^\ast - a\right)/\sqrt{2}$ &$U$  \\
    $\left(b^\ast-b, a + \ii a^\ast, a^\ast + \ii a, -b^\ast - b\right)/\sqrt{2}$ &$X\,U$ \\
    $\left(-\ii b -\ii b^\ast, a^\ast + \ii a, -\ii a^\ast -  a, -\ii b + \ii b^\ast\right)/\sqrt{2}$ &$Y\,U$ \\
    $\left(a^\ast-a, b^\ast + \ii b, b +\ii b^\ast, a + a^\ast \right)/\sqrt{2}$ &$Z\,U$ \\
\hline
\end{tabular}
\label{tb:1qubitgate}
\end{table}

We recall that our model requires a single measurement on a four-level physical site to implement any arbitrary single-qubit gate, 
which is in contrast to the one-way model wherein a general single-qubit gate can be implemented with four measurements on two-level sites.
In addition, it is evident that if our operation gives a by-product $E$, which is a local unitary, we can simply try to perform $E^\dagger$ factor at the next stage. 
i.e., if we want to apply one-qubit gate $U$, we apply $UE^\dagger$ instead.
This procedure removes the by-product up to this stage. 
\subsection{Two-qubit gate}

To realize a two-qubit gate, we need a two-dimensional structure. We assign the following rank-3 tensors to circle sites:

\begin{align}\label{def:ucrcl}
    T(0)&=\ket{\phi(0)} \otimes B(0), &
    T(1)&=\ket{\phi(1)} \otimes B(1), \nonumber \\
    T(2)&=\ket{\phi(2)} \otimes B(2), &
    T(3)&=\ket{\phi(3)} \otimes B(3),
\end{align}
where
\begin{equation}\label{def:B}
B = \left( \openone, X, Z, ZX \right)
\end{equation} 
and $ \ket{\phi(0)} = \ket{\phi(1)} = \ket{0}, \ket{\phi(2)} = \ket{\phi(3)} = \ket{1}$.

\begin{figure}
  \centering
  \includegraphics[height=40mm]{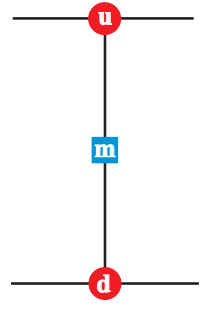}\\
  \caption{Sketch of performing a two-qubit gate: measure all the three sites.}\label{fig:biqubitgate}
\end{figure}

In the following, we propose a scheme for the construction of the two-qubit CZ gate.
According to the model, in each vertical edge we have three sites (Fig.~\ref{fig:biqubitgate}). 
We first measure the middle square site such that a Hadamard gate is realized in the correlation space as a result.
Next we perform suitable measurements---as explained later in Eq.~\eqref{def:CZbase}---on the both sides of this square site such that the overall result becomes tantamount to implementing a CZ gate (up to a Pauli by-product). We remark that order is not important in the above procedure.

Let $m$, $u$, and $d$ be the outcomes of the measurements on the middle, upper, and lower sites (Fig.~\ref{fig:biqubitgate}), respectively. Using  Eq.~(\ref{eq:2qubitgate}), one can show
\begin{align} \label{eq:biqubitgate}
W(d,m,u) = &\sum_{s,t} \psi^\ast_d(s) \psi^\ast_u(t) \,c(s,t;m)\, B(s) \otimes
B(t), \nonumber \\
& c(s,t;m):=\bra{\phi(s)}E_\mathrm{sq}(m)\,H\ket{\phi(t)}.
\end{align}
Here $H$ is the Hadamard gate, and can be implemented following the rules in Table~\ref{tb:1qubitgate}. Now, we measure both `u' and `d' sites in the following basis:
\begin{align}\label{def:CZbase}
0:&\;(1,0,1,0)/\sqrt{2}, &
1:&\;(0,1,0,-1)/\sqrt{2}, \nonumber \\
2:&\;(1,0,-1,0)/\sqrt{2}, &
3:&\;(0,1,0,1)/\sqrt{2},
\end{align}
where the labels of the vectors indicate the corresponding measurement results. The resulting gate is 
\begin{multline}\label{eq:ECZE}
    W(d,m,u)= \\
    E_{\mathrm{m}}(m)\;E_\mathrm{ l}(d)\!\otimes\! E_\mathrm{ l}(u)\;\mbox{\large CZ}\;
    E_\mathrm{ r}(d)\!\otimes\! E_\mathrm{ r}(u)\;E_{\mathrm{m}}(m),
\end{multline}
in which
\begin{align}\label{eq:Eset}
E_{\mathrm{m}}=&\; (\openone\otimes \openone, \openone\otimes X, X \otimes \openone, X\otimes X),
\\
E_\mathrm{ l}=&\;(\openone,X,X,\openone),\quad
E_\mathrm{ r}=\;(\openone,\openone,X,X).
\end{align}
As an example, consider the ideal case where all the three measurement outcomes are 0. The table of $c(s,t;0)$ is
\begin{center}
\begin{tabular}{c|cccc}
\hline s/t & 0 & 1 & 2 & 3 \\ 
\hline 0 & 1 & 1 & 1 & 1 \\ 
 1 & 1 & 1 & 1 & 1 \\ 
 2 & 1 & 1 & -1 & -1 \\ 
 3 & 1 & 1 & -1 & -1 \\ 
\hline 
\end{tabular}
\end{center}
Thus Eq.~\eqref{eq:biqubitgate} yields
\begin{equation}
W(0,0,0)= \frac{1}{2} \big( \openone \otimes \openone + \openone \otimes Z + Z \otimes \openone - Z \otimes Z \big),
\end{equation}
which is the expansion of the CZ gate at the Pauli basis. 
As can be seen in Eqs.~\eqref{eq:ECZE} and \eqref{eq:Eset}, all possible by-products are uncorrelated Pauli by-products.

\subsection{Removal of an unwanted vertical edge}

\begin{figure}
\centering
  \includegraphics[height=35mm]{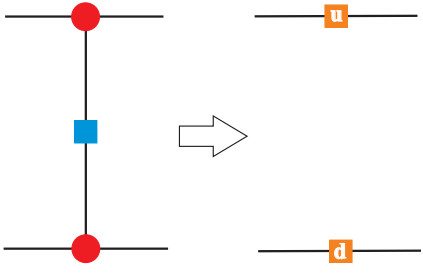}\\
  \caption{Removal of a vertical edge by measuring corresponding mid-site.}\label{fig:removal}
\end{figure}

Assume that we have arrived at a circle site but do not aim to apply a two-qubit gate thereon.
Thus we need to remove the corresponding vertical edge by suitable measurements on the associated mid (square) site.
If the resulting operation in the correlation space is in the form of some $\ket{\beta}_\mathrm{ d}\bra{\alpha}_\mathrm{ u}$, then the vertical edge is removed thereby.
It is further favorable to transform controllably the left disjoint circle sites to some other sites used in computation
(see Fig.~\ref{fig:removal}).
All these requirements can be fulfilled if, for example, we have: $\alpha = \beta = +$, because by contracting $\bra{+}$ with the tensors (\ref{def:ucrcl}), the resulting list of matrices become $B$ [Eq.~(\ref{def:B})], which is a universal list of matrices for computation.
In fact, this choice is a realization of $\ketbra{+}{+}$ operation on the mid site.
Consider the following measurement basis and the resulting operators:
\begin{align}\label{def:baseofremoval}
    &0:\ (2,1+\ii,1+\ii,0)/2\sqrt{2} &\rightarrow & &\ketbra{+}{+},\\
    &1:\ (0,\ii-1,1-\ii,2)/2\sqrt{2} &\rightarrow & &\ketbra{+}{-},\\
    &2:\ (0,1-\ii,\ii-1,2)/2\sqrt{2} &\rightarrow & &\ketbra{-}{+},\\
    &3:\ (2,-1-\ii,-1-\ii,0)/2\sqrt{2} &\rightarrow & &\ketbra{-}{-}.
\end{align}
Depending on the outcome of the measurement, a by-product occurs on the
new up and down squares. 
The rule to find these by-products is straightforward: whenever either of 
$\beta$ or $\alpha$ becomes ``$-$'' in $\ket{\beta}_\mathrm{d}\bra{\alpha}_\mathrm{u}$, then an $X$ by-product
sandwiches the matrices of the corresponding site. For example, for
outcome `2', we have
\begin{align}\label{eq:err_removal}
B_\mathrm{ u}(s) &= B(s), &
B_\mathrm{ d}(s) &= X\,B(s)\,X.
\end{align}

\subsection{Readout of the result}
\begin{figure}
\centering
 \includegraphics[width=40mm]{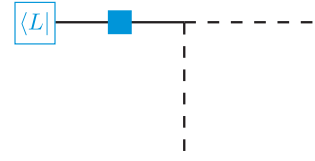}\\
  \caption{The square site is the last site of the row. We want to get information about
  the state of the qubit on that site in the correlation space by measuring this physical site.}\label{fig:readout}
\end{figure}
At the end of the computation process to read the results we proceed as follows.
 Suppose we reached the last site of a row (Fig.~\ref{fig:readout}).
In our model, we choose the left boundary as $\bra{L}=\bra{0}$. 
From Eqs.~\eqref{eq:readout} and \eqref{eq:model-A}, we have the following $\bra{L(i)}$ vectors: 
\begin{align}\label{eq:L(i)}
    \bra{L(0)}&=\tfrac{1}{\sqrt{2}}\bra{0}, &  
    \bra{L(1)}&=\tfrac{1}{\sqrt{2}}\bra{1}, \nonumber \\
    \bra{L(2)}&=\tfrac{\ii}{\sqrt{2}}\bra{1}, &
    \bra{L(3)}&=\tfrac{1}{\sqrt{2}}\bra{0}.
\end{align}
As a result, if in the correlation space, the state associated to a final (i.e., readout) site is
\begin{equation}
\ket{\psi}:=\psi_0\ket{0}+\psi_1\ket{1}
\end{equation}
its corresponding physical state from Eq.~(\ref{eq:readout-operator}) becomes
 \begin{equation}
\ket{\Psi}=\frac{1}{\sqrt{2}}\left(\psi_0\ket{0}+\psi_1\ket{1}+\ii \psi_1\ket{2}+\psi_0\ket{3}\right).
 \end{equation}
Now it can be seen that the following projective measurement on the corresponding physical site:
\begin{align}
0:\;&\ketbra{0}{0}+\ketbra{3}{3},\\
1:\;&\ketbra{1}{1}+\ketbra{2}{2},
\end{align}
gives rise to an equivalent $Z$-measurement on the correlation space.

\subsection{A parent Hamiltonian}\label{sec:parham}
Here we follow the general recipe of Refs.~\cite{cirac06,perez08} to construct parent Hamiltonians for MPS and PEPS.
The brief sketch of this construction is comprised of the following three steps:
\begin{enumerate}
\item Partition the lattice into regions including (sufficiently large) neighboring sites.
\item Calculate the associated reduced density matrix of each pair of neighboring regions.
\item Write the parent Hamiltonian as the summation of the projection onto the null space of each density matrix.
\end{enumerate}
According to Ref.~\cite{perez08}, in step 1 if for each region the associated list of tensors forms a complete basis, then the PEPS is the \emph{unique} ground state of the constructed Hamiltonian.
This is why the region should comprises \emph{sufficiently large} number of sites to produce a complete basis (if possible).
 \begin{figure}
\centering
  \includegraphics[width=35mm]{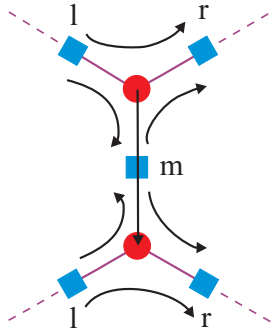}\\
  \caption{A honeycomb lattice for the 3-local Hamiltonian of our resource PEPS. Here to each site, two virtual sites are attributed.}\label{fig:h}
\end{figure}
 
 Following the above sketch and after some (tedious) calculations, we found a three-local Hamiltonian whose unique ground state is the very resource state we have introduced in Sec.~\ref{sec:model}; see Appendix~\ref{app:ham}.
In our calculations, we have partitioned the lattice into two types of regions: any circle site with its right neighboring square site is a region, in addition, each vertical square site is another region.
  It is straightforward to see that the associated list of tensors of each of these regions forms a complete basis.

\section{Summary}
We have analyzed some conditions which lead to a deterministic universal measurement-based quantum computation in `correlation' space of projected entangled-pair states (PEPS). 
These conditions are essentially related to the propagation of by-products produced during the computation due to the randomness of the measurement outcomes. 
In particular, we have obtained some sufficient conditions for the simulation of two-quDit gates and readout of the final results on the correlation space.

In particular, we proposed a qubit measurement-based model on a four-level honeycomb lattice which satisfies all the  conditions we derived in this paper.
 The model is a PEPS on a four-level honeycomb lattice which is used for qubit quantum computation. 
 One of the properties of the model is that a general single-qubit gate can be implemented by only one single-site measurement. 
 In addition, all the by-products produced during the computation are local Pauli operators.
 It has also been shown that the PEPS we used is a unique ground state of a three-local Hamiltonian.
 
An interesting next step to pursue is to look for new universal PEPSs with perhaps more desirable properties.
Additionally, since how the by-products propagate within the model was essential to its structure, we anticipate that perhaps employing fault-tolerant schemes borrowed from the circuit model can be used to improve the class of models based on PEPSs.

\section*{Acknowledgments}
We appreciate illuminating discussions with V. Karimipour and A. T. Rezakhani. 
We also thank the anonymous referees for valuable comments which improved the quality and presentation of our results.

\bibliographystyle{apsrev}
\bibliography{refs}

\appendix

\section{Local unitary operators crossing two-qubit gates locally} \label{seq:err-two-qubit}

Any arbitrary two-qubit gate can be written as
\begin{equation}
\ee^{\ii \theta}\,k_1\, \ee^{  \frac{\ii }{2} \left( \alpha\, X\otimes X + \beta\, Y \otimes Y + \gamma\, Z \otimes Z \right) } \, k_2,
\end{equation}
where $\alpha,\beta,\gamma \in [0,\pi)$ and $k_1,\, k_2 \in \mathrm{SU}(2)\otimes \mathrm{SU}(2)$ are local gates \cite{Zh03}. 
Considering the non-trivial part
\begin{equation}\label{eq:W-gate}
W :=  \ee^{ \frac{\ii }{2} \left( \alpha\, X\otimes X + \beta\, Y \otimes Y + \gamma\, Z \otimes Z \right) }
\end{equation}
one may look for local unitaries $u,\,u' \in \mathrm{U}(2) \otimes \mathrm{U}(2)$ such that
\begin{equation}\label{eq:wuw}
u' = W u W^\dagger.
\end{equation}
It is straightforward to check that these $u$'s constitute a subgroup of $ \mathrm{U}(2) \otimes \mathrm{U}(2)$.
Thus the question reduces to finding the generators of this subgroup.


The key idea to solve this problem is to use the isomorphism $ \mathrm{SU}(2) \otimes \mathrm{SU}(2) \cong \mathrm{SO}(4)$.
To see this, let
\begin{equation}
Q= \frac{1}{\sqrt{2}} \begin{bmatrix}
1 & 0 & 0 & \ii \\
0 & \ii & 1 & 0  \\
0 & \ii & -1 & 0 \\
1 & 0 & 0 & - \ii 
\end{bmatrix},
\end{equation}
from whence for any $k \in \mathrm{SU}(2) \otimes \mathrm{SU}(2)$ we have
$Q^\dagger k Q \in \mathrm{SO}(4)$ (see, e.g., Ref.~\cite{Zh03}). Hence the problem becomes to find all $O \in \mathrm{SO}(4)$ and $\eta \in [0,2\pi)$ such that the following condition is satisfied 
\begin{equation}
O' :=\ee^{-\ii \eta} D O D^\dagger \in \mathrm{SO}(4),
\end{equation}
where $D=Q^\dagger W Q$.
Here we have replaced Eq.~\eqref{eq:wuw} with its equivalent form $\ee^{\ii \eta} k' = W k W^\dagger,$ in which $k,\, k' \in \mathrm{SU}(2)\otimes \mathrm{SU}(2)$ and $\eta$ is a real number.
Note that \cite{Zh03}
\begin{multline}
D = \\ \mathrm{diag}\left(  \ee^{\frac{\ii}{2}\left( \alpha - \beta + \gamma \right) },\ee^{\frac{\ii}{2}\left( \alpha + \beta - \gamma \right) },\ee^{\frac{\ii}{2}\left( -\alpha - \beta - \gamma \right) },\ee^{\frac{\ii}{2}\left( -\alpha + \beta + \gamma \right) }\right) .
\end{multline}

Orthogonality of $O'$ implies
\begin{equation}
 D\, O {D^\ast}^2 \trnsp{O} D = \ee^{2 \ii \eta}
\end{equation}
or equivalently
\begin{equation}
D^2 O {D^2}^\ast = \ee^{2 \ii \eta} O.
\end{equation}
This equetion in turn can be rewritten in the following equivalent form:
\begin{equation}\label{eq:filter-O}
F \ast O = \ee^{2 \ii \eta }\,O,
\end{equation}
where 
\begin{equation}
F = \begin{bmatrix}
1 & \ee^{2 \ii \left(- \beta + \gamma \right) } & \ee^{2 \ii \left(\alpha + \gamma \right) } & \ee^{2 \ii \left( \alpha - \beta \right) } \\
\ee^{2 \ii \left( \beta - \gamma \right) } & 1 & \ee^{2 \ii \left( \alpha + \beta \right) } & \ee^{2 \ii \left( \alpha - \gamma \right) } \\
\ee^{2 \ii \left( -\alpha -\gamma \right) } & \ee^{2 \ii \left( -\alpha - \beta \right) } & 1 & \ee^{2 \ii \left( - \beta - \gamma \right) } \\
\ee^{2 \ii \left( -\alpha + \beta \right) } & \ee^{2 \ii \left( -\alpha + \gamma \right) } & \ee^{2 \ii \left( \beta + \gamma \right) } & 1
\end{bmatrix},
\end{equation}
and $\ast$ is the element-wise multiplication operator, defined through
\begin{equation}
(F\ast O)_{ij} := F_{ij} O_{ij}.
\end{equation}

Let the pair $\langle O_\eta,\,\eta \rangle$ be a solution of Eq.~\eqref{eq:filter-O}. Then one can infer that
\begin{equation}
\ee^{-2 \ii \eta } F_{ij} \neq 1\ \Rightarrow \ (O_\eta)_{ij} = 0.
\end{equation}

One can always uniquely decompose $F$ as
\begin{equation} \label{eq:filter-Fmu}
F=\sum_{\eta \in \mathrm{Solution}} \ee^{2\ii \eta}F_\eta,
\end{equation}
where $\eta \in [0,\pi)$ and $F_\eta$ are non-zero binary matrices, in the sense that $(F_\eta)_{ij}\in \{0,1\}$.
Hence one can write a collection of equations equivalent to Eq.~\eqref{eq:filter-O}:
\begin{equation} \label{eq:filter-O-Fmu}
F_\eta \ast O_\eta = O_\eta,
\end{equation}
where $F_\eta$ determines the only possible elements of $O_\eta$ that are non-zero.
Let us define a map $\mathcal{N}$ that identifies non-zero elements of any matrix $M$, i.e.,
\begin{equation}
(\mathcal{N}[M])_{ij}=\left\lbrace \begin{matrix}
1 & M_{ij} \neq 0 \\
0 & M_{ij} = 0
\end{matrix}\right. .
\end{equation}
Using this operation, Eq.~\eqref{eq:filter-O-Fmu} can be read as
\begin{equation}\label{eq:Omu-Fmu}
\mathcal{N}[O_\eta] = F_\eta.
\end{equation}

In this part, we find solutions to Eq.~\eqref{eq:Omu-Fmu} for  seven special cases,
from which six cases correspond to the generators of SO(4) group,
while the other case deals  diagonal even reflections (i.e., diagonal SO(4) matrices).
These diagonal matrices can only be either of the eight matrices in the form diag$\left(\pm 1,\pm 1, \pm 1, \pm 1 \right)$ with even number of $-1$'s.
\\

\begin{itemize}
\item[``0'')]

In the decomposition \eqref{eq:filter-Fmu} one of the $F_\eta$'s may be $\openone$.
For this case, $O_0$ should be one of the diagonal SO(4) matrices (as stated above).
This solution in turn implies the following list of local gates:
\begin{equation}
L^{(0)} = \left( \openone \otimes \openone,\; X \otimes X,\; Y \otimes Y,\; Z \otimes Z \right),
\end{equation}
all of which obviously commute with the two-qubit gate $W$ of Eq.~\eqref{eq:W-gate}.

\item[``12'')]
If $O_\eta$ is a rotation in 12-plane, i.e.,
 \begin{equation}
 O_\eta= \begin{bmatrix}
 \cos \theta & - \sin \theta & & \\
 \sin \theta & \cos \theta & & \\
 & & 1 & \\
 & & & 1
 \end{bmatrix}
 \end{equation}
(up to some diagonal even reflections), the corresponding $F_\eta$ is obtained as
\begin{equation}
F_\eta=\begin{bmatrix}
1 & 1 & & \\
1 & 1 & & \\
  &   & 1 & \\
  &  &  & 1
\end{bmatrix}.
\end{equation}
This $O_\eta$ is equivalent to $k= Q O_\eta Q^\dagger= X(\theta) \otimes X(\theta)$, 
where $X(\theta) \equiv \mathrm{exp}( \ii\, X\,  \theta / 2)$ is the rotation about the x-axis by the angle $\theta$ in spin-1/2 space.
Thus the list of the solution for this case is
 \begin{equation}
 L^{(12)}(\theta) = X(\theta) \otimes X(\theta),
 \end{equation}
up to any multiplicative factor from $L^{(0)}$. 

The remaining five other cases are obtained in a similar way.
In the following we only list the solution local unitary gate (again up to multiplicative factor from $L^{(0)}$).

\item[``13'')] If $O_\eta$ is a rotation in the 13-plane, we obtain
\begin{align}
F_\eta=&\begin{bmatrix}
1 &  & 1& \\
 & 1 & & \\
1&   & 1 & \\
  &  &  & 1
\end{bmatrix}, \\
L^{(13)}(\theta)=& Y(\theta) \otimes Y(-\theta).
\end{align}

\item[``14'')] If $O_\eta$ is a rotation in the 14-plane, then
\begin{align}
F_\eta=& \begin{bmatrix}
1 &  & &1 \\
 & 1 & & \\
  &   & 1 & \\
 1 &  &  & 1
\end{bmatrix}, \\
L^{(14)}(\theta)=& Z(\theta) \otimes Z(\theta).
\end{align}

\item [``23'')] If $O_\eta$ is a rotation in the 23-plane, we obtain
\begin{align}
F_\eta=& \begin{bmatrix}
1 &  & & \\
 & 1 &1 & \\
  &  1 & 1 & \\
  &  &  & 1
\end{bmatrix}, \\
L^{(23)}(\theta)=& Z(-\theta) \otimes Z(\theta).
\end{align}

\item[``24'')] If $O_\eta$ is a rotation in the 24-plane, then
\begin{align}
F_\eta=&\begin{bmatrix}
1 &  & & \\
 & 1 & & 1\\
  &   & 1 & \\
  & 1 &  & 1
\end{bmatrix}, \\
L^{(24)}(\theta)=& Y(\theta) \otimes Y(\theta).
\end{align}

\item[``34'')] If $O_\eta$ is a rotation in the 34-plane, we obtain
\begin{align}
F_\eta=& \begin{bmatrix}
1 &  & & \\
 & 1 & & \\
  &   & 1 &1 \\
  &  & 1 & 1
\end{bmatrix}, \\
L^{(34)}(\theta)=& X(-\theta) \otimes X(\theta).
\end{align}
\end{itemize}

\emph{Example}: Consider the CZ gate\footnote{Note also that the CNOT gate is also locally equivalent to CZ.}, which is locally equivalent to $ \mathrm{exp}(\ii \frac{\pi}{4} Z \otimes Z)$ as follows:
\begin{equation}
\mathrm{CZ} = \ee^{\ii \frac{\pi}{4}}\, Z(-\pi/2)\otimes Z(-\pi/2) \, \mathrm{exp}(\ii \frac{\pi}{4}\, Z \otimes Z).
\end{equation}
Thus $\gamma=\pi/2,\,\alpha=\beta=0$, and $F$ is
 \begin{equation}
 F= \begin{bmatrix}
 1 & -1 & -1 & 1\\
 -1 & 1 & 1 & -1\\
 -1 & 1 & 1 & -1\\
 1 & -1 & -1 & 1
 \end{bmatrix}.
 \end{equation}
 The decomposition of this $F$ gives rise to
 \begin{equation}
 F = F_0 - F_{\pi/2},
 \end{equation}
 where
 \begin{align}
   F_0 =& \begin{bmatrix}
   1 &  &  & 1\\
    & 1 & 1 & \\
    & 1 & 1 & \\
   1 &  &  & 1
   \end{bmatrix}, \\
   F_{\pi/2}=& \begin{bmatrix}
   & 1 & 1 &  \\
   1& & &1 \\
   1 & & & 1\\
   & 1 & 1 &
   \end{bmatrix}.
 \end{align}
$F_0$ includes the 14- and the 23-plane rotations. 
Here the corresponding local gates are constituted from products of matrices from the list $L^{(14)},\,L^{(23)}$, and $L^{(0)}$:
\begin{align}
 L_0(t_1,t_2,i) &= L^{(14)}(t_1) L^{(23)}(t_2) L^{(0)}(i) \nonumber \\
  & = Z(t_1) \otimes Z(t_1)\, Z(-t_2)\otimes Z(t_2)\,L^{(0)}(i) \nonumber \\
  & = Z(t_1-t_2) \otimes Z(t_1+t_2)\, L^{(0)}(i).
\end{align}
This equation can also be simplified as
\begin{equation}\label{eq:L0}
L_0(\theta_1,\theta_2,i) = Z(\theta_1) \otimes Z(\theta_2)\, L^{(0)}(i),
\end{equation}
for any $\theta_1, \theta_2$, and $0\leq i < 4$.
Without lose of generality, this gives the form of the general form of the $L_0$ family of local gates 
which pass through the CZ gate.

A remark here is in order. 
Note that the matrix $F_{\pi/2}$ can not be written as a sumation over the $F_\eta$'s obtained in the seven special cases studied earlier.
However one can see that $O_{\pi/2}=O_0 T$, where 
\begin{equation}
T = \begin{bmatrix}
  & 1 &  &  \\
 1 &  &  &  \\
  &  &  & 1 \\
  &  & 1 & 
\end{bmatrix} \in \mathrm{SO(4)}.
\end{equation}
That is
\begin{equation}
\mathcal{N}[O_0]=F_0\ \iff \ \mathcal{N}[O_0 T ] = F_{\pi/2}. 
\end{equation}
Since the local gate equivalent to $T$ is $Y \otimes Z$, 
the family $L_{\pi/2}$, which represents all local gates corresponding the $\eta=\pi/2$ case, 
can be written as
\begin{equation}
L_{\pi/2}(\theta_3,\theta_4,j) = L_0(\theta_3,\theta_4,j)\,Y\otimes Z. 
\end{equation}

Composition of $L_0$ and $L_{\pi/2}$  makes the group $L$ which is the group of all local gates that pass through the CZ gate locally (in the sense Eq.~\eqref{eq:wuw}). 
The $L$ family can be parametrize as follows by combining $L_0$ and $L_{\pi/2}$ together:
\begin{equation}\label{eq:local-cross-CZ}
L(\theta_1,\theta_2,i,j)= Z(\theta_1) \Sigma(i) \otimes Z(\theta_2)\Sigma(j) ,
\end{equation}
where $\Sigma = \left( \openone,\, X,\, Y,\, Z\right)$. 

It is important to note that the above solution is sensitive to the value of the $\gamma$ parameter.
For example, consider the case $\gamma \neq \pi/2,\,\alpha=\beta=0$, where $F$ is
\begin{equation}
\begin{bmatrix}
1 & \ee^{2 \ii \gamma } & \ee^{2 \ii \gamma } & 1 \\
 \ee^{-2 \ii \gamma } & 1 & 1 & \ee^{-2 \ii \gamma } \\
 \ee^{-2 \ii \gamma } & 1 & 1 & \ee^{-2 \ii \gamma } \\
 1 & \ee^{2 \ii \gamma } & \ee^{2 \ii \gamma } & 1
\end{bmatrix} = F_0 +  \ee^{2 \ii \gamma } F_{\gamma} + \ee^{-2 \ii \gamma } F_{\pi-\gamma}, 
\end{equation}
with
\begin{equation}
F_{\gamma} = \begin{bmatrix}
 & 1 & 1 &  \\
  &  &  &  \\
  &  &  &  \\
  & 1 & 1 & 
\end{bmatrix},\ 
F_{\pi-\gamma} = \begin{bmatrix}
 &  &  &  \\
 1 &  &  & 1 \\
 1 &  &  & 1 \\
  &  &  & 
\end{bmatrix}.
\end{equation}
There is no solution for Eq.~\eqref{eq:filter-O-Fmu} for the above $F_\gamma,\, F_{\pi-\gamma}$.
This means that there is not any special orthogonal matrix $O$ such that $\mathcal{N}[O]=F_{\gamma}$ or $\mathcal{N}[O]=F_{\pi - \gamma}$. 
Thus for this family of two-qubit gates, the group of local gates passing locally through them is only the $L_0$ of Eq.~\eqref{eq:L0}.

\section{Parent Hamiltonian of the example model}\label{app:ham}
Following the sketch briefly described in Sec.~\ref{sec:parham} (and more extensively discussed in Ref.~\cite{cirac06}), we now explicitly construct a 3-local parent Hamiltonian for our resource PEPS.

Consider the set of basis operators
\begin{align}\label{def:base}
    \Sigma^{\otimes 2}=& \left(\Sigma(i)\otimes \Sigma(j)\:|\: 0 \leq i,j < 4 \right), \\
    \Sigma :=& (\openone,X,Y,Z),
\end{align}
where a 4-level site is taken as two 2-level ``virtual'' sites. In the following, we have used a shorthand which can be understood by this example: $0122(-1)+3200(2)$ denotes $-\openone\otimes X\ \text{{\large $\otimes$}}\ Y \otimes Y + 2 \,Z \otimes Y\ \text{{\large $\otimes$}}\ \openone \otimes \openone$.

Note that due to non-commutativity of the terms of the Hamiltonian in the Pauli-group basis, the Hamiltonian is not a stabilizer Hamiltonian.

Details of the derivation of the Hamiltonian are analytically cumbersome, thus here we report the final result obtained by numerical programming (Fig.~\ref{fig:h}):

\begin{align}\label{eq:model-ham}
H = \sum_{\text{vertical edges}} & h^{\mathrm{u}}_{\mathrm{lr}} + h_{\mathrm{lr}}^\mathrm{d} + h_{\mathrm{lum}} +  h_{\mathrm{ldm}} + \nonumber \\ & h_{\mathrm{mur}} + h_{\mathrm{mdr}} +  h_{\mathrm{umd}},
\end{align}
\begin{align}
&h_{\mathrm{lr}}^{\mathrm{u/d}}  = \nonumber \\ &
000000(28)+000101(-2)+000102(-2)+000113(-2)+\nonumber \\ & 
000123(-2)+103001(-1)+103002(-1)+103013(-1)+\nonumber \\ & 
103023(-1)+103100(-2)+103211(-1)+103212(1)+\nonumber \\ & 
103221(-1)+103222(1)+103301(1)+103302(1)+\nonumber \\ & 
103313(-1)+103323(-1)+110201(-1)+110202(-1)+\nonumber \\ & 
110213(1)+110223(1)+110311(-1)+110312(1)+\nonumber \\ & 
110321(-1)+110322(1)+120201(-1)+120202(-1)+\nonumber \\ & 
120213(1)+120223(1)+120311(-1)+120312(1)+\nonumber \\ & 
120321(-1)+120322(1)+203001(-1)+203002(-1)+\nonumber \\ & 
203013(-1)+203023(-1)+203100(-2)+203211(-1)+\nonumber \\ & 
203212(1)+203221(-1)+203222(1)+203301(1)+\nonumber \\ & 
203302(1)+203313(-1)+203323(-1)+210201(1)+\nonumber \\ & 
210202(1)+210213(-1)+210223(-1)+210311(1)+\nonumber \\ & 
210312(-1)+210321(1)+210322(-1)+220201(1)+\nonumber \\ & 
220202(1)+220213(-1)+220223(-1)+220311(1)+\nonumber \\ & 
220312(-1)+220321(1)+220322(-1)+313001(-1)+\nonumber \\ & 
313002(-1)+313013(-1)+313023(-1)+313100(-2)+\nonumber \\ & 
313211(1)+313212(-1)+313221(1)+313222(-1)+\nonumber \\ & 
313301(-1)+313302(-1)+313313(1)+313323(1)+\nonumber \\ & 
323001(-1)+323002(-1)+323013(-1)+323023(-1)+\nonumber \\ & 
323100(-2)+323211(1)+323212(-1)+323221(1)+\nonumber \\ & 
323222(-1)+323301(-1)+323302(-1)+323313(1)+\nonumber \\ & 
323323(1), 
\end{align}

\begin{align}
& h_\mathrm{lum} = \nonumber \\ &
000000(28)+003011(-2)+003012(-2)+003021(2)+\nonumber \\ & 
003022(2)+100111(-1)+100112(-1)+100121(1)+\nonumber \\ & 
100122(1)+101110(-1)+101120(-1)+101131(1)+\nonumber \\ & 
101132(1)+102110(1)+102120(1)+102131(1)+\nonumber \\ & 
102132(1)+103100(-2)+111010(-1)+111020(-1)+\nonumber \\ & 
111031(-1)+111032(-1)+112010(-1)+112020(-1)+\nonumber \\ & 
112031(1)+112032(1)+121010(-1)+121020(-1)+\nonumber \\ & 
121031(-1)+121032(-1)+122010(-1)+122020(-1)+\nonumber \\ & 
122031(1)+122032(1)+200111(-1)+200112(-1)+\nonumber \\ & 
200121(1)+200122(1)+201110(-1)+201120(-1)+\nonumber \\ & 
201131(1)+201132(1)+202110(1)+202120(1)+\nonumber \\ & 
202131(1)+202132(1)+203100(-2)+211010(1)+\nonumber \\ & 
211020(1)+211031(1)+211032(1)+212010(1)+\nonumber \\ & 
212020(1)+212031(-1)+212032(-1)+221010(1)+\nonumber \\ & 
221020(1)+221031(1)+221032(1)+222010(1)+\nonumber \\ & 
222020(1)+222031(-1)+222032(-1)+310111(-1)+\nonumber \\ & 
310112(-1)+310121(1)+310122(1)+311110(1)+\nonumber \\ & 
311120(1)+311131(-1)+311132(-1)+312110(-1)+\nonumber \\ & 
312120(-1)+312131(-1)+312132(-1)+313100(-2)+\nonumber \\ & 
320111(-1)+320112(-1)+320121(1)+320122(1)+\nonumber \\ & 
321110(1)+321120(1)+321131(-1)+321132(-1)+\nonumber \\ & 
322110(-1)+322120(-1)+322131(-1)+322132(-1)+\nonumber \\ & 
323100(-2),
\end{align}

\begin{align}
& h_\mathrm{ldm} = \nonumber \\ &
000000(28)+003011(-2)+003012(2)+003021(-2)+\nonumber \\ & 
003022(2)+100111(-1)+100112(1)+100121(-1)+\nonumber \\ & 
100122(1)+101101(-1)+101102(-1)+101113(1)+\nonumber \\ & 
101123(1)+102101(1)+102102(1)+102113(1)+\nonumber \\ & 
102123(1)+103100(-2)+111001(-1)+111002(-1)+\nonumber \\ & 
111013(-1)+111023(-1)+112001(-1)+112002(-1)+\nonumber \\ & 
112013(1)+112023(1)+121001(-1)+121002(-1)+\nonumber \\ & 
121013(-1)+121023(-1)+122001(-1)+122002(-1)+\nonumber \\ & 
122013(1)+122023(1)+200111(-1)+200112(1)+\nonumber \\ & 
200121(-1)+200122(1)+201101(-1)+201102(-1)+\nonumber \\ & 
201113(1)+201123(1)+202101(1)+202102(1)+\nonumber \\ & 
202113(1)+202123(1)+203100(-2)+211001(1)+\nonumber \\ & 
211002(1)+211013(1)+211023(1)+212001(1)+\nonumber \\ & 
212002(1)+212013(-1)+212023(-1)+221001(1)+\nonumber \\ & 
221002(1)+221013(1)+221023(1)+222001(1)+\nonumber \\ & 
222002(1)+222013(-1)+222023(-1)+310111(-1)+\nonumber \\ & 
310112(1)+310121(-1)+310122(1)+311101(1)+\nonumber \\ & 
311102(1)+311113(-1)+311123(-1)+312101(-1)+\nonumber \\ & 
312102(-1)+312113(-1)+312123(-1)+313100(-2)+\nonumber \\ & 
320111(-1)+320112(1)+320121(-1)+320122(1)+\nonumber \\ & 
321101(1)+321102(1)+321113(-1)+321123(-1)+\nonumber \\ & 
322101(-1)+322102(-1)+322113(-1)+322123(-1)+\nonumber \\ & 
323100(-2),
\end{align}

\begin{align}
& h_\mathrm{mur} = \nonumber \\ &
000000(28)+000101(-2)+000102(-2)+000113(-2)+\nonumber\\&
000123(-2)+101201(-1)+101202(-1)+101213(1)+\nonumber\\&
101223(1)+101311(-1)+101312(1)+101321(-1)+\nonumber\\&
101322(1)+102201(-1)+102202(-1)+102213(1)+\nonumber\\&
102223(1)+102311(-1)+102312(1)+102321(-1)+\nonumber\\&
102322(1)+113000(-2)+113101(-1)+113102(-1)+\nonumber\\&
113113(-1)+113123(-1)+123000(-2)+123101(-1)+\nonumber\\&
123102(-1)+123113(-1)+123123(-1)+201201(-1)+\nonumber\\&
201202(-1)+201213(1)+201223(1)+201311(-1)+\nonumber\\&
201312(1)+201321(-1)+201322(1)+202201(-1)+\nonumber\\&
202202(-1)+202213(1)+202223(1)+202311(-1)+\nonumber\\&
202312(1)+202321(-1)+202322(1)+213000(2)+\nonumber\\&
213101(1)+213102(1)+213113(1)+213123(1)+\nonumber\\&
223000(2)+223101(1)+223102(1)+223113(1)+\nonumber\\&
223123(1)+311201(-1)+311202(-1)+311213(1)+\nonumber\\&
311223(1)+311311(-1)+311312(1)+311321(-1)+\nonumber\\&
311322(1)+312201(1)+312202(1)+312213(-1)+\nonumber\\&
312223(-1)+312311(1)+312312(-1)+312321(1)+\nonumber\\&
312322(-1)+321201(-1)+321202(-1)+321213(1)+\nonumber\\&
321223(1)+321311(-1)+321312(1)+321321(-1)+\nonumber\\&
321322(1)+322201(1)+322202(1)+322213(-1)+\nonumber\\&
322223(-1)+322311(1)+322312(-1)+322321(1)+\nonumber\\&
322322(-1),
\end{align}

\begin{align}
& h_\mathrm{mdr} = \nonumber \\ &
000000(28)+000101(-2)+000102(-2)+000113(-2)+\nonumber \\ & 
000123(-2)+011201(-1)+011202(-1)+011213(1)+\nonumber \\ & 
011223(1)+011311(-1)+011312(1)+011321(-1)+\nonumber \\ & 
011322(1)+012201(-1)+012202(-1)+012213(1)+\nonumber \\ & 
012223(1)+012311(-1)+012312(1)+012321(-1)+\nonumber \\ & 
012322(1)+021201(-1)+021202(-1)+021213(1)+\nonumber \\ & 
021223(1)+021311(-1)+021312(1)+021321(-1)+\nonumber \\ & 
021322(1)+022201(-1)+022202(-1)+022213(1)+\nonumber \\ & 
022223(1)+022311(-1)+022312(1)+022321(-1)+\nonumber \\ & 
022322(1)+113000(-2)+113101(-1)+113102(-1)+\nonumber \\ & 
113113(-1)+113123(-1)+123000(2)+123101(1)+\nonumber \\ & 
123102(1)+123113(1)+123123(1)+131201(-1)+\nonumber \\ & 
131202(-1)+131213(1)+131223(1)+131311(-1)+\nonumber \\ & 
131312(1)+131321(-1)+131322(1)+132201(1)+\nonumber \\ & 
132202(1)+132213(-1)+132223(-1)+132311(1)+\nonumber \\ & 
132312(-1)+132321(1)+132322(-1)+213000(-2)+\nonumber \\ & 
213101(-1)+213102(-1)+213113(-1)+213123(-1)+\nonumber \\ & 
223000(2)+223101(1)+223102(1)+223113(1)+\nonumber \\ & 
223123(1)+231201(-1)+231202(-1)+231213(1)+\nonumber \\ & 
231223(1)+231311(-1)+231312(1)+231321(-1)+\nonumber \\ & 
231322(1)+232201(1)+232202(1)+232213(-1)+\nonumber \\ & 
232223(-1)+232311(1)+232312(-1)+232321(1)+\nonumber \\ & 
232322(-1),
\end{align}

\begin{align}
& h_\mathrm{umd} = \nonumber \\  &
000000(6)+001130(-1)+001230(1)+002130(-1)\nonumber \\ & 
+002230(1)+301100(-1)+301200(-1)+302100(1)\nonumber \\ & 
+302200(1)+303330(-2).\label{eq:model-ham-stb}
\end{align}

\end{document}